\documentclass{elsart4-1}

\usepackage{graphicx}
\usepackage{amssymb}

\usepackage[english,francais]{babel}

\newtheorem{e-proposition}[theorem]{Proposition}

\newtheorem{e-definition}[theorem]{Definition\rm}

\renewcommand{\theequation}{\arabic{equation}}
\def\og{\leavevmode\raise.3ex\hbox{$\scriptscriptstyle\langle\!\langle$~}}
\def\fg{\leavevmode\raise.3ex\hbox{~$\!\scriptscriptstyle\,\rangle\!\rangle$}}

\begin{document}

\begin{frontmatter}


\selectlanguage{english}
\title{Expected Performance of a \\ Self-Coherent Camera}


\selectlanguage{english}
\author[adress:obspm]{Raphael Galicher},
\ead{raphael.galicher@obspm.fr}
\author[adress:obspm]{Pierre Baudoz}
\ead{pierre.baudoz@obspm.fr}

\address[adress:obspm]{\small Observatoire de Paris-Meudon\\
5 place Jules Janssen, 92195 Meudon, France}

\begin{abstract}
Residual wavefront errors in optical elements limit the performance of coronagraphs. To improve their efficiency, different types of devices have been proposed to correct or calibrate these errors. In this paper, we study one of these techniques proposed by Baudoz et al. 2006~\cite{Baudoz06} and called Self-Coherent Camera~(SCC). The principle of this instrument is based on the lack of coherence between the stellar light and the planet that is searched for. After recalling the principle of the SCC, we simulate its performance under realistic conditions and compare it with the performance of differential imaging.

{\it To cite this article:  R.~Galicher, P.~Baudoz, C.~R.~Physique~8 (2007), 333-339}

\vskip 0.5\baselineskip

\selectlanguage{francais}
\noindent{\bf R\'esum\'e}
\vskip 0.5\baselineskip
\noindent
{\bf \'Etude des Performances d'une Self-Coherent Camera.}
La qualit\'e de surface des optiques limite les performances des coronographes. La correction ou l'\'{e}talonnage de ces d\'efauts optiques permet d'en am\'eliorer l'efficacit\'e. Nous \'etudions dans cet article une technique, propos\'ee par Baudoz et al. 2006~\cite{Baudoz06}, qui permet d'\'{e}talonner les tavelures cr\'e\'ees au plan focal par les d\'efauts de surface d'onde. Le principe de cet instrument, appel\'e Self-Coherent Camera (SCC), est bas\'e sur l'absence de coh\'erence entre l'\'etoile et la plan\`ete. Apr\`es un rappel du principe de la SCC, nous pr\'esentons une comparaison de ses performances avec celles de l'imagerie diff\'erentielle.

{\it Pour citer cet article~:  R.~Galicher, P.~Baudoz, C.~R.~Physique~8 (2007), 333-339}

 \vskip 0.5\baselineskip
\noindent{\small{\it Mots-cl\'es~:} Imagerie \`a Haut Contraste~; Plan\`ete Extrasolaire~; Traitement des images}}
\end{abstract}
\end{frontmatter}


\selectlanguage{english}
\section{Introduction}
\label{}
     Radial velocimetry has enabled the indirect detection of more than 200 exoplanets over the past decade~\cite{Schneider07}. The study of their physical parameters involves the measure of their spectra and a straightforward solution is their direct detection. However, even the  brightest planets~\cite{Baraffe03} are 10$^7$ to 10$^9$ times fainter than their host and often located within a fraction of an arcsecond of the star. A large number of coronagraphs have been proposed to suppress the host star's overwhelming flux~\cite{Guyon06} but all of them are limited by the imperfections of the wavefront. Using high-order Adaptive Optics (AO) or spatial observatory, the performance of coronagraphs is still limited by the aberrations in the coronagraphic optics and the residual errors from the AO (for ground-based observation).  
Thus, when attempting to detect a faint companion on long exposure images, residual speckle patterns can be a dominant source of error~\cite{Marois05}. Since these speckle patterns are slowly drifting, it is mandatory to find a way to discriminate the speckles of the star from a faint companion during the exposure. This is the purpose of differential imaging techniques. Several criteria have been proposed so far to discriminate the speckles from the planets: spectrophotometry~\cite{Marois05,Racine99}, polarimetry~\cite{Seager00,Baba03}, and coherence~\cite{Baudoz06,Codona04,Guyon04,Labeyrie04}. While both concepts based on spectrophotometry and polarimetry depends on the physical properties of the planets, the coherence is a robust criterion when no physical information is available from the planets that could be observed. Here, we study the way to calibrate the speckles using coherence as proposed by Baudoz et al. 2006~\cite{Baudoz06}. First, we recall the concept of the technique, called Self Coherent Camera (SCC). Then, we compare the detectability of a companion using the SCC with classical techniques of differential imaging.

\section{Principle}
\label{principle}
     
 The concept of the SCC has already been presented~\cite{Baudoz06} but the main features are recalled here. The purpose of that device is to discriminate in a field of view a companion image from the speckles created by wavefront defects. While both features look almost the same on the detector, only the speckles are coherent with the stellar beam. A Fizeau recombination is used to encode the field of view with a coherent fringe pattern that affects only the stellar speckles. The principle and a possible set-up are described in figure~\ref{fig:principe}. The light coming from the telescope is split into two beams. One of the beams is spatially filtered using a pinhole or an optical fiber. The typical size of the pinhole is about the size of the Point Spread Function core ($\approx \lambda/D$) for two reasons: 1) Since high frequency pupil defects are diluted in the focal plane, the pinhole cleans the wavefront of the reference beam \cite{Ollivier97}. 2) The companion is not transmitted whenever its distance to the star is larger than the size of the pinhole. The two beams are recombined in the focal plane in a Fizeau scheme. To do so, the two pupil beams are optically brought at the same plane right before a focusing lens. Because of the Fizeau interferences, the intensity distribution of the stellar flux is fringed. Because of the spatial filtering, the pupil illumination of the reference beam is not anymore uniform. Thus, the fringe contrast will not be 100\%. The contrast will also be diminished by differential aberrations between the reference and the main beam. However, section~\ref{Formalism} and~\ref{sec : simu} show that it does not limit the SCC performance. Since the flux of the companion is removed from the reference beam by spatial filtering, the intensity of the companion will be unaltered by the reference beam and the image of the companion will not be fringed (figure~\ref{fig:principe}). 

\section{Formalism}
\label{Formalism}
The electromagnetic field in the entrance pupil plane is described by $\Psi(\xi)=\widehat{A}(\xi)$. The term $A(x)$ is the complex amplitude in the focal plane and $\widehat{A}(\xi)$ denotes the Fourier Transform of $A(x)$. Coordinates $\xi$ and $x$ are used for the pupil plane and the focal plane respectively. The field of both the star and its companion in the pupil plane can be written :$\Psi_*(\xi)+\Psi_C(\xi)=\widehat{A_*}(\xi)+\widehat{A_C}(\xi)$. The coherent pupil of the reference beam can be described in the pupil plane by:
$\Psi_R(\xi)=\widehat{A_R}(\xi)$

\begin{figure}[ht]
\begin{center}
\includegraphics[width=14cm]{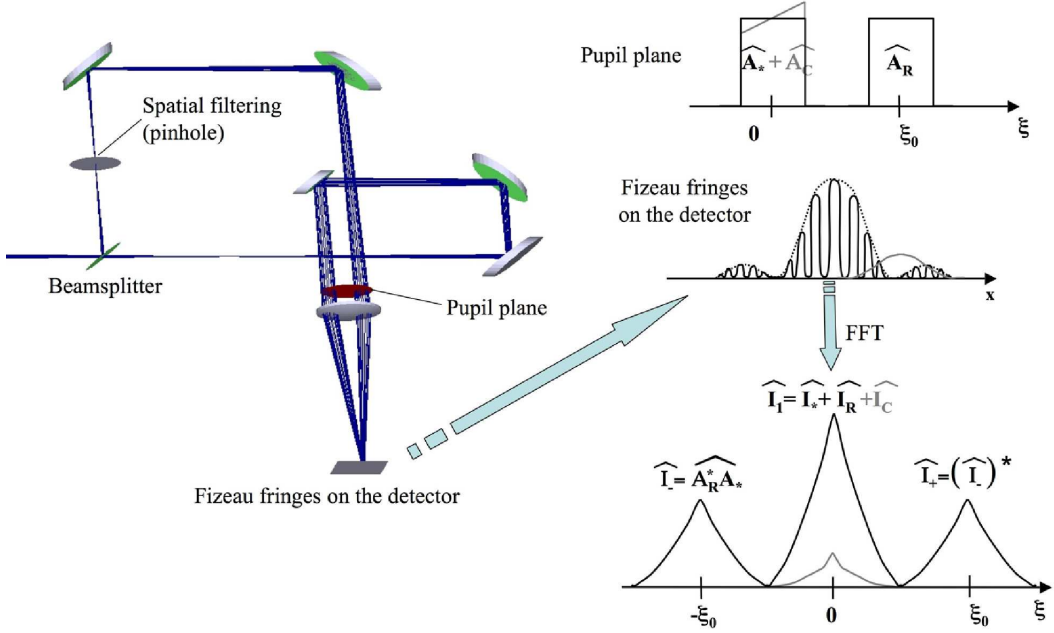}
\caption[]{Left: Possible Set-up for the SCC. Right: Principle of the SCC. The electromagnetic field distribution in the SCC pupil is indicated for the star  and its reference beam (black) and a companion (grey). At the bottom right, the numerical Fourier Transform of the detected image shows autocorrelations and intercorrelations between the reference beam and the stellar beam. \label{fig:principe}}
\end{center}
\end{figure}

Assuming that the vector ${\xi_0}$ describes the distance between the coherent pupil and the corrugated pupil (supposed to be centred on zero),  the field at the SCC pupil output plane is given by:
\begin{equation} 
\Psi_{SCC}(\xi)=\Psi_*(\xi)+\Psi_C(\xi)+\Psi_R(\xi)*\delta(\xi- \xi _0)
\end{equation}
 where $*$ is the convolution symbol. For a given complex amplitude $A(x)$, we write the intensity detected at the focal plane as $I(x)=\mid A(x) \mid^2$.
Assuming the optical path difference of the interferometer is kept at zero, the image recorded at the focal plane is:

\begin{center}
  \begin{equation}
    I(x)=\mid\widehat{\Psi_{SCC}}\mid^2 = I_*(x)+I_C(x)+I_R(x)+2 Real \{A_*(x) A^*_R(x)\}\cdot\cos(2\pi x \xi_0)
    \label{eq:image}
  \end{equation}
\end{center}

The image looks like a classical Fizeau image with fringes pinning the intensity distribution of the main star. However, the image of the companion does not show any fringes because it is not coherent with the reference beam.
Since the intensity of the main star is coded with fringes while the image of the companion is not, it looks clearly that one can discriminate the image of the companion from the stellar flux. The simplest approach is to use the FFT of the image to separate the three areas limited by the autocorrelation of the pupil function. The centred area is the sum of the autocorrelations of the three images~$I_R$, $I_*$ and~$I_C$. The correlation between the fields~$\Psi_*$ and $\Psi_R$ appears in lateral areas~(Fig.~\ref{fig:principe}). The separation between the two images of the pupil ${\xi _0}$ is large enough so that the different terms are not superimposed and can be numerically separated.

Assuming that we describe the centred area by $\hat{I}_1=\hat{I}_*+\hat{I}_C+\hat{I}_R$ and the two other areas by $\hat{I}_-=\widehat{A^*_R . A_*}$ and $\hat{I}_+=(\hat{I}_-)^*$, we can write the intensity of the companion using the following formula $\displaystyle I_C=I_1-I_R-\frac{I_+.I_-}{I_R}$.

To detect the companion $I_C$, one needs to record separately $I_R$. It can be done because $I_R$ is a spatially filtered beam that can be very stable over time. In fact, the stability of $I_R$ is a possible limitation of the SCC. A laboratory experiment is under development to analyze the impact of $I_R$ instability. Assuming we record $\overline{I_R}$, which is an estimate of $I_R$, the final $I_C$ estimator, called~$SCC$, is given by\,:

\begin{equation}
SCC=I_1-\overline{I_R}-\frac{I_+.I_-}{\overline{I_R}}\label{eq : SCCIR}
\end{equation}

This equation is valid only for an exposure shorter than the coherent time of the atmospheric speckles and for monochromatic beam~\cite{Baudoz06}. However, using a chromatic corrector as proposed by Wynne 1979~\cite{Wynne79}, the bandwidth could be increased up to a reasonable value (R=5). 
Equation~\ref{eq : SCCIR} is true for any aberration or illumination of $I_*$ or $I_R$. Thus, neither the non-uniform illumination of $I_R$ because of spatial filtering nor any differential aberration between the reference and the main beam should limit the detection of a companion with the SCC.

\section{Numerical simulations}
\label{sec : simu}
To confirm that result, we developed a numerical code that simulates the SCC. In this section, we decribe the hypothesis we assumed and detail the data processing of the simulation. The main physical hypothesis and parameters are listed underneath\,:
\begin{itemize}
\item The telescope is 8-meter diameter.
\item The considered star is a 5th visible magnitude G star observed at $0.8\mu m$ with $R=\frac{\lambda}{\Delta\lambda}=8$ assuming a perfect Wynne corrector is used\cite{Wynne79}.
\item The quantum efficiency of the camera is set to $0.4$. The exposure time must be shorter than the coherence time, so we choosed an exposure time of~$6\,ms$.  As shown by Sarazin and Tokovinin's study\,\cite{Sa01}, the atmospheric coherence time at the wavelength of $0.8\,\mu m$ is longer than $6\,ms$ more than~$50\%$ of the time at Paranal.
\item We used the approach proposed by Rigaut et al. 1998~\cite{Ri98,Jo02} to generate the atmospheric phase screens of the Paranal site. Shack-Hartmann wavefront sensor with $80$ actuators across the telescope diameter and a $1\,ms$ AO closed loop temporal have been assumed. The seeing has been set to~$0.6\,arcsec$ at $0.8\,\mu m$. A global AO Power Spectral Density (PSD) taking into account servo-lag, aliasing and fitting errors has been computed. From this DSP, independent phase screens realizations are created.
\item Following Cavarroc et al. 2006~\cite{Ca06}, we introduced common~$\delta_C$ and non-common~$\delta_{NC}$ static aberrations. For the simulated SCC (respec.~differential imaging) device, $\delta_C$ are static aberrations in the instrument upstream of the SCC~(respec.~differential imaging instrument). $\delta_{NC}$~are the differential static phase aberrations between the corrugated and the reference beams for both techniques. $\delta_C$ is set to~$10nm$ rms and $\delta_{NC}$ to~$5nm$ or~$1nm$ rms and both  are created from a PSD following a $f^{-2}$ law\,\cite{Du02}.
\item Photon noise, but no read-out-noise, is taken into consideration.
\item Images are $512$x$512$ pixels and the entrance pupil diameter is $D=80$ pixels. The simulation is monochromatic.
\end{itemize}

\subsection{Self-Coherent Camera}
To simulate the SCC we split the telescope AO beam into two beams (see section~\ref{principle}) which are both corrugated by the AO corrected aberrations and~$\delta_C$. The reference beam is filtered by a $\frac{\lambda}{D}$~diameter pinhole in a focal plane. Then, an entrance pupil size diaphragm is used to stop the diffracted light into the following pupil plane. Finally, that reference pupil is corrugated by a non-common static phase aberrations,~$\delta_{NC}$. Then, the two beams, separated by~$2.15\,D$, are recombined in a Fizeau scheme assuming a zero optical path difference. The resulting focal plane intensity is recorded in a numerical image. 

The SCC data processing is done in three steps. The first one is a Fast Fourier Transform of the interferometric image. Then, the three autocorrelation areas, $\hat{I}_1$,~$\hat{I}_+$~and~$\hat{I}_-$, are separated using masks and an inverse Fast Fourier Transform gives~$I_1$, $I_+$ and $I_-$. Finally, the SCC residual image is given by equation\,\ref{eq : SCCIR}, where $\overline{I_R}$ is a long exposure recording of the reference image. However, the division by~$\overline{I_R}$ leads to undefined values where $\overline{I_R}$ equals zero. Thus, the image~$SCC.\overline{I_R}$ is computed instead of the $SCC$ one. Afterwards, we divide the resulting image by~$\overline{I_R}$ using a~$10^{-2}$~threshold. \subsection{Differential imaging device}
 Both beams of that device are corrugated by the AO corrected aberrations and~$\delta_C$. One of the beams, called here the reference beam for SCC comparison, is also corrugated by~$\delta_{NC}$. The differential image is the subtraction of the reference image to the other one.
The simulation of differential imaging is optimistic because both beams are supposed observed at the same wavelength and at the same polarization. Moreover, the companion is introduced in only one image and completely removed from the other beam.

\subsection{Detectability}
\label{subsec : detectability}
The goal of the two previous simulated devices is to detect a faint companion. Calling~$C$ the companion energy contrast, the image maximum intensity is~$C\,I_{*max}$, where~$I_{*max}$ is the host star image maximum intensity. To determine which contrast can be detected by the considered device, we define~$D$, the detectability at~$5\sigma$, as $D = 5\frac{N}{I_{*max}}$, where~$N$ is the residual noise into the final image when no companion is present. In that paper, $N$~is computed as the square root of the azimutal spatial variance of the final image.

\section{Results of the simulation}
\subsection{Detectability versus angular separation}
\begin{figure}
\begin{center}
\includegraphics[width=6.8 cm]{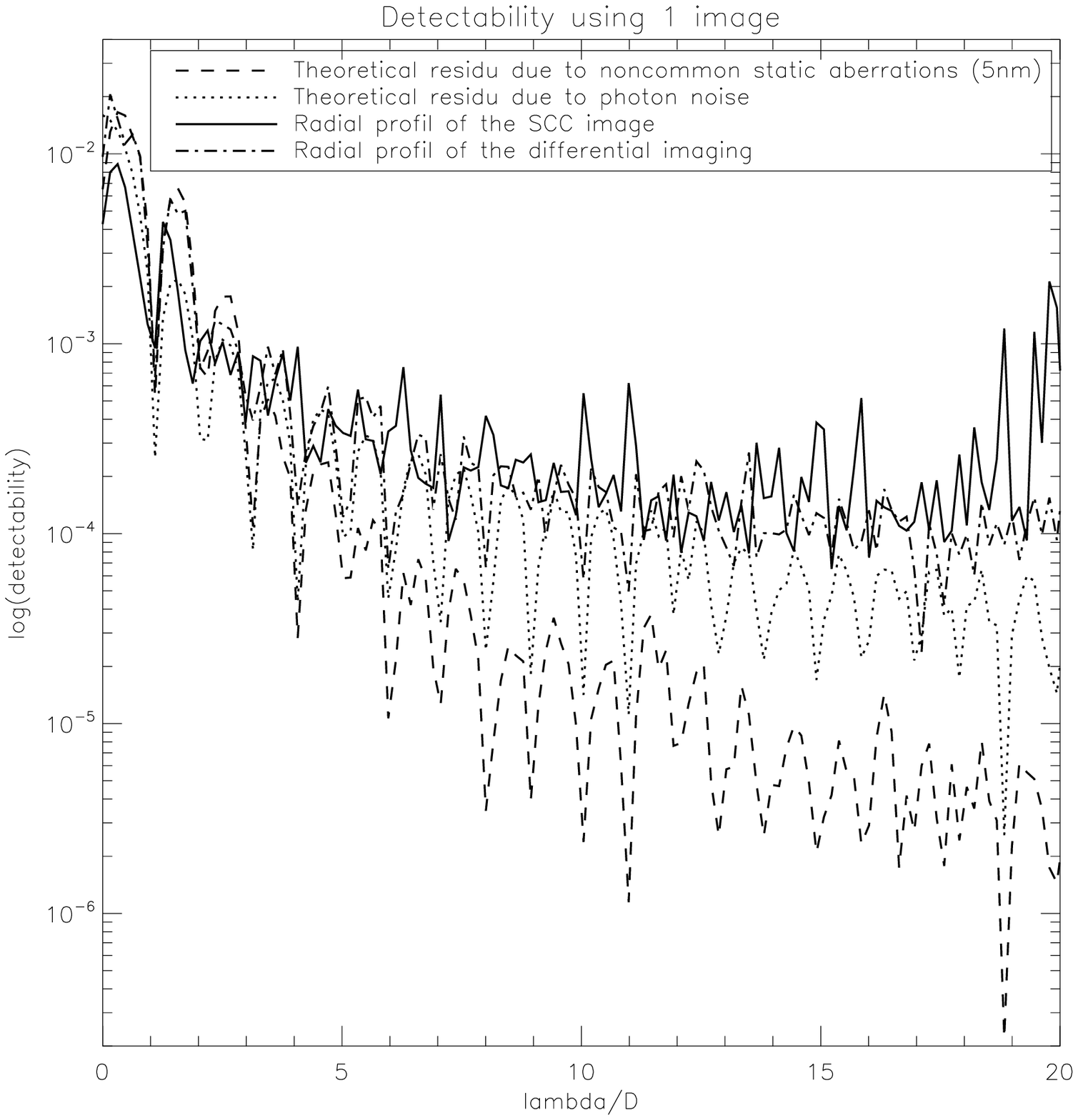}
\includegraphics[width=6.8 cm]{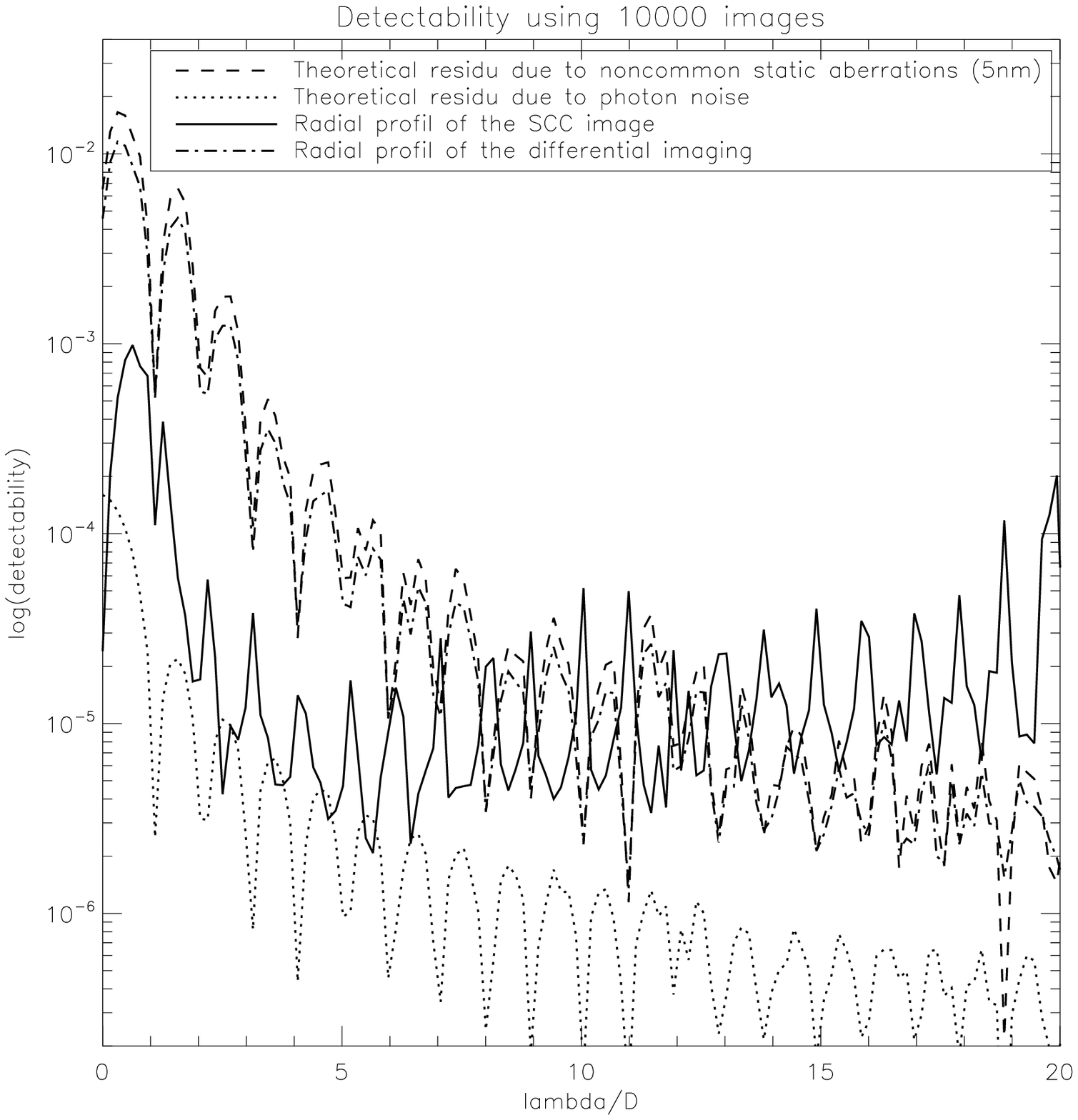}
\end{center}
\caption{\it SCC~(solid line) and differential imaging~(dot-dashed line) detectabilities versus angular separation . Theoretical photon noise~(dotted line) and~$5\,nm$~rms non-common static aberrations~(dashed line) are also plotted. The SCC reference image exposure time is $18\,s$. Left: One single image of~$6\,ms$. Right: 10000 images of~$6\,ms$.}
\label{fig : detect_rad}
\end{figure}
Figure~\ref{fig : detect_rad} shows SCC~(solid line) and differential imaging~(dot-dashed line) detectabilities at~$5\sigma$ versus the angular separation to the host star. For the SCC, the~$\overline{I_R}$ exposure time is~$18\,s$ to minimize the photon noise. Using a single image (Figure~\ref{fig : detect_rad}~left), the theoretical photon noise~(dotted line) is greater than the~$5\,nm$~rms non-common static aberrations~(dashed line). The SCC and differential imaging detectabilities roughly follow the theoretical photon noise limit. However, the SCC profile shows spikes and an amplification for angular separation greater than $8\frac{\lambda}{D}$. These spikes and amplification are due to the~$\overline{I_R}$ division where~$\overline{I_R}$ is smaller than one photon.
For a sum of 10000 images, the theoretical photon noise is decreased by a factor 100 (figure~\ref{fig : detect_rad}, right). Thus, the non-common static aberrations start to dominate in the differential image as predicted by Cavarroc et al. 2006\,\cite{Ca06}. The SCC noise almost follows the photon noise limitation and is not limited by static aberrations where the reference image~$\overline{I_R}$ is greater than one photon. However, spikes are greater than on the single image profile~(figure~\ref{fig : detect_rad}, left). Other data processing are under study to minimize these spikes. 

\subsection{Detectability versus number of used images}\label{subsec : detect_nbim}
\begin{figure}
\begin{center}
\includegraphics[width=11 cm]{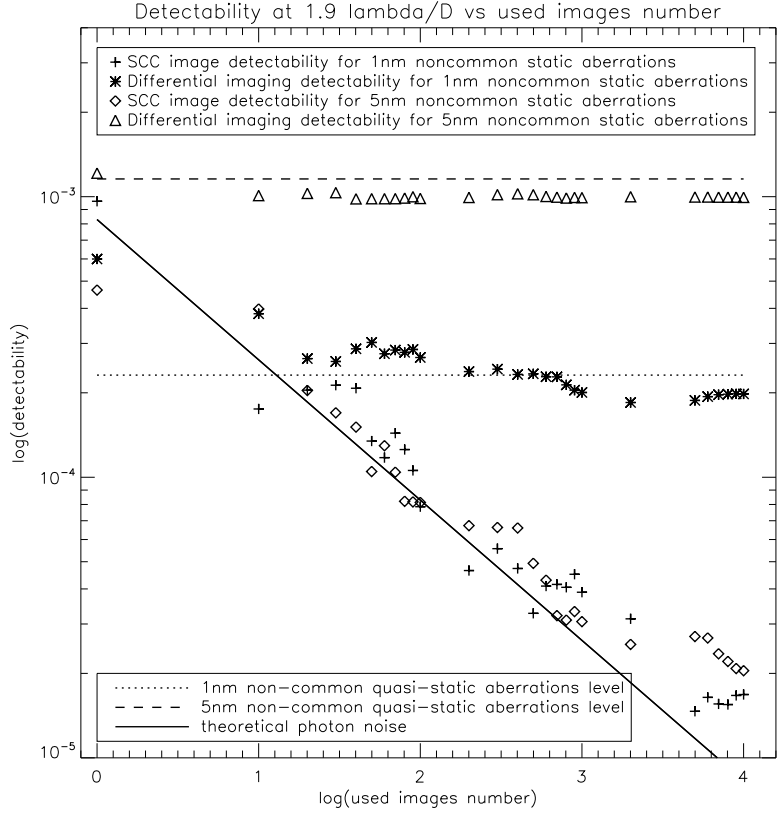}
\end{center}
\caption{\it SCC~(plus and diamonds) and differential imaging~(crosses and triangles) detectabilities at~$5\sigma$ at~$1.9\frac{\lambda}{D}$ versus the number of used images for~$1\,nm$ and~$5\,nm$~rms non-common static aberrations~(respec. dotted and dashed line). Photon noise~(solid line) at~$1.9\frac{\lambda}{D}$ is also plotted. Log-log scale is used. The SCC reference image exposure time is $18\,s$.}
\label{fig : detect_nbim}
\end{figure}
To check that the SCC noise decreases as the square root of the number of images recorded (following the photon noise), we plotted on figure~\ref{fig : detect_nbim} the SCC detectability~(plus and diamonds) at $1.9\frac{\lambda}{D}$ versus the number of used images. In comparison, we added the same plot for the differential imaging detectability~(crosses and triangles). We also plotted theoretical photon noise~(solid line) and non-common static aberrations noises~(dotted and dashed lines for respectively $1\,nm$ and~$5\,nm$~rms). As seen in figure~\ref{fig : detect_rad}, the differential imaging, unlike the SCC, is limited by non common static aberrations. Furthermore, where~$\overline{I_R}$ is greater than one photon, the SCC image is limited by photon noise and decreases as the square root of the number of used images. However, there is a minimum detectability that the Self-Coherent Camera can reach. Indeed, even if the photon noise is minimized on~$\overline{I_R}$ using an $18\,s$~exposure time, it is still present and limits the SCC detectability~(flat evolution after~$5000$~images). The exact impact of the reference beam noise is still under study.

\subsection{Resulting images}
\begin{figure}
\begin{center}
\includegraphics[width=12 cm]{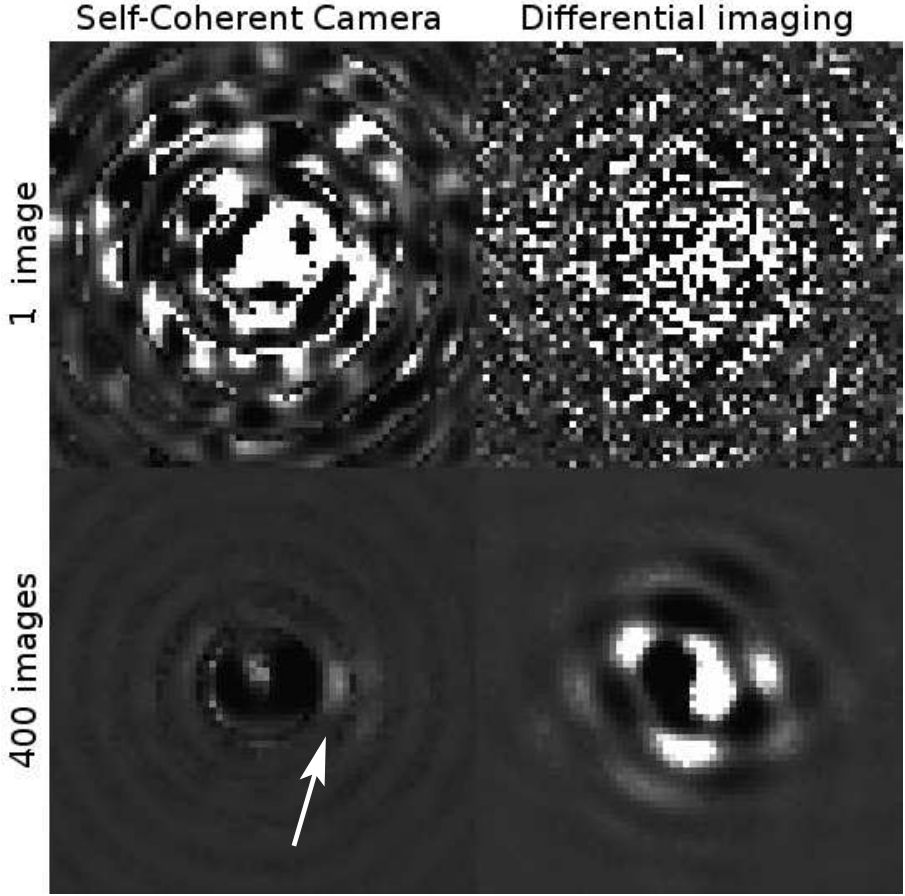}
\end{center}
\caption{\it Final Self-Coherent Camera~(left) and differential~(right) images using a single~(top) and $400$~images~(bottom). The companion, $1.9\frac{\lambda}{D}$~separated, is $10^{4}$~less bright than the host star. The SCC reference image exposure time is $18s$. An algorithm based on cosmic removal is used to correct the SCC images from spikes due to the division by $\overline{I_R}$. Power scale is the same for the four images.}
\label{fig : comp_images}
\end{figure}
From figure~\ref{fig : comp_images}, it looks like $400$~SCC images (corresponding to only~$2.4\,s$) are sufficient to detect a $10^{-4}$~companion at an angular separation of~$1.9\frac{\lambda}{D}$ from the host star. This is true whatever the static aberration amplitude is. On the contrary, the differential imaging technique is limited by non-common static aberrations. We simulated a companion-star system and observed the corresponding images (figure~\ref{fig : comp_images}). As expected, the companion is not detected on the final images using a single image for both techniques that are photon noise limited~(top of the figure). Using $400$~images, the companion is lost in non-common static speckles on the final differential image whereas it is well detected on the final SCC image.

\section{Conclusion}
In this paper, we have recalled the principle of the Self-Coherent Camera (SCC). We have described the numerical simulation we developed to evaluate the performance of the SCC. We found that, as opposed to standard differential imaging, the SCC is not limited by the aberrations in the optical elements. The detection limit with the SCC roughly follow the photon noise limitation~(cf.~figure~\ref{fig : detect_rad} and~\ref{fig : detect_nbim}). Thus, to detect a faint companion very close to its host star~($10^{-4}$ at~$1.9\frac{\lambda}{D}$), only a short observation time is mandatory (about $2.4\,s$ for a magnitude 5 star on an 8-meter telescope). To detect fainter companion as a $10^{-10}$~exoplanet, the coupling of coronagraphs and the SCC will be necessary. This coupling is under study. We are also analyzing different data processing technique to minimize the impact of the division by the reference image~$\overline{I_R}$. A laboratory experiment is also under development to compare the expected capabilities presented in this paper with the effective performance.




\end{document}